\RequirePackage{fix-cm}                 

\documentclass{article}

\usepackage[english]{babel}             
\usepackage[T1]{fontenc}                

\PassOptionsToPackage{hyphens}{url}

\usepackage[
  hidelinks,                            
  pdfusetitle,                          
  pdftitle={%
    Learned Compression of Encoding Distributions%
  },                                    
  colorlinks,                           
  linkcolor=black,                      
]{hyperref}

\usepackage{iftex}                      

\usepackage{amssymb}
\usepackage{booktabs}                   
\usepackage{caption}
\usepackage[inline]{enumitem}           
\usepackage{fancyhdr}                   
\usepackage{float}                      
\usepackage{geometry}
\usepackage{graphicx}
\usepackage{makecell}                   
\usepackage{mathtools}
\usepackage{microtype}                  
\usepackage{multirow}                   
\usepackage{textcomp}                   
\usepackage{xcolor}                     
\usepackage{xfrac}                      

\usepackage{csquotes}

\MakeOuterQuote{"}

\usepackage{alphalph}

\usepackage[
  capitalize,
  nameinlink,
]{cleveref}

\crefname{equation}{}{}

\usepackage{spconf}

\title{Learned Compression of Encoding Distributions}
\name{Mateen Ulhaq and Ivan V. Baji\'c}
\author{Mateen Ulhaq \and Ivan V. Baji\'c}
\address{%
  School of Engineering Science,
  Simon Fraser University,
  Burnaby, BC, Canada%
}

\newif\ifarxiv
\arxivtrue

\ifarxiv
  \fancypagestyle{firstpage}{
    \renewcommand{\topmargin}{-8mm}
    \setlength{\headsep}{2.0\normalbaselineskip}

    \fancyhead{}
    \fancyfoot{}
    \fancyhead[C]{%
      \footnotesize%
      \textit{Proc. IEEE ICIP 2024}, Abu Dhabi, UAE, Oct. 2024 \\
      \vspace{0.6\normalbaselineskip}%
      \tiny%
      \copyright{} 2024 IEEE. Personal use of this material is permitted. Permission from IEEE must be obtained for all other uses, in any current or future media, including reprinting/republishing this material for advertising or promotional purposes, creating new collective works, for resale or redistribution to servers or lists, or reuse of any copyrighted component of this work in other works.%
    }
  }
\fi

\DeclareMathOperator{\E}{\mathbb{E}}

\DeclareMathOperator*{\argmin}{arg\,min}
\newcommand{\boldvar}[1]{{\boldsymbol{#1}}}

\ifx\sfrac\undefined
  \providecommand{\nicefrac}[2]{{#1}/{#2}}
\else
  \providecommand{\nicefrac}[2]{\sfrac{#1}{#2}}
\fi
\providecommand{\sfrac}[2]{\nicefrac{#1}{#2}}

\newenvironment{subsubfigure}[2][]{%
  \begin{subfigure}[#1]{#2}%
    \stepcounter{subsubfigure}%
}{%
    \addtocounter{subfigure}{-1}%
  \end{subfigure}%
}
\newcounter{subsubfigure}

\newlength{\tablecaptionbelowskip}
\newlength{\tablecaptionaboveskip}
\begin{document}

\maketitle

\ifarxiv
  \thispagestyle{firstpage}
\fi

\begin{abstract}
  The entropy bottleneck introduced by Ballé~\emph{et~al.}~\cite{balle2018variational} is a common component used in many learned compression models.
  It encodes a transformed latent representation using a static distribution whose parameters are learned during training.
  However, the actual distribution of the latent data may vary wildly across different inputs.
  The static distribution attempts to encompass all possible input distributions, thus fitting none of them particularly well.
  This unfortunate phenomenon, sometimes known as the amortization gap, results in suboptimal compression.
  To address this issue, we propose a method that dynamically adapts the encoding distribution to match the latent data distribution for a specific input.
  First, our model estimates a better encoding distribution for a given input.
  This distribution is then compressed and transmitted as an additional side-information bitstream.
  Finally, the decoder reconstructs the encoding distribution and uses it to decompress the corresponding latent data.
  Our method achieves a Bjøntegaard-Delta (BD)-rate gain of \textminus{}7.10\% on the Kodak test dataset when applied to the standard fully-factorized architecture.
  Furthermore, considering computational complexity, the transform used by our method is an order of magnitude cheaper in terms of Multiply-Accumulate (MAC) operations compared to related side-information methods such as the scale hyperprior.
\end{abstract}


\section{Introduction}
\label{sec:pdf_compression/intro}



\begin{figure}[tbp]
  \centering
  \includegraphics[width=1.0\linewidth]{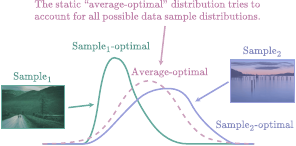}
  \caption[Suboptimality of static amortized encoding distributions]{%
    Visualization of the suboptimality of using a single static encoding distribution.
    This distribution is optimal, on average, among all static distributions, but it is suboptimal for any specific data instance.%
  }
  \label{fig:pdf/amortization-gap}
\end{figure}



In this paper, we explore the compression of probability distributions that are used in compression.
Specifically, we will focus on probability distributions that are used to encode the latent representations produced by an image compression model.
Entropy models compress latents using \emph{encoding \mbox{distributions}},
which are discrete probability mass functions (pmfs) used for encoding.
For example, Ballé~\emph{et~al.}~\cite{balle2018variational} encode a latent representation $\boldvar{y}$ using the following entropy models:
\begin{enumerate}
  \item
    A "fully-factorized" \emph{entropy bottleneck}.
    Each element ${y}_{c,i}$ within the $c$-th channel is encoded using a single channel-specific non-parametric encoding distribution ${p}_{c}(y)$, which is typically fixed once training completes.
  \item
    A \emph{Gaussian conditional}.
    For each element ${y}_{i}$, the parameters ${\mu}_{i}$ and ${\sigma}_{i}$ of its encoding distribution are computed, conditioned on the latent $\boldvar{\hat{z}} = \operatorname{Quantize}[h_a(\boldvar{y})]$.
    In later works~\cite{cheng2020learned}, a Gaussian mixture model (GMM) is sometimes used instead,
    with parameters ${w}_{i}^{(k)}$, ${\mu}_{i}^{(k)}$, and ${\sigma}_{i}^{(k)}$ for each Gaussian in the mixture.
\end{enumerate}
Of these two, the entropy bottleneck is more flexible at modeling arbitrary distributions.
However, because all elements within a given channel are modelled using the same distribution, it is quite poor at specializing and adapting to specific input distributions.
Instead, it models a "conservative" distribution that tries to encompass all possible inputs, which is often suboptimal, as visualized in \cref{fig:pdf/amortization-gap}.
This is sometimes known as the \emph{amortization gap}~\cite{balcilar2022amortizationgap,cremer2018inferencesuboptimality}.
In contrast, the Gaussian conditional is better at adapting to a specific input, though it is limited to modeling Gaussian distributions.
Furthermore, it must also pay an additional cost for transmitting the distribution parameters, though this is often worth the trade-off, e.g., a savings of roughly 40\% for a 10\% increase in rate in~\cite{balle2018variational}.

In this paper, we propose a method to address this amortization gap via the compression of an input-specific encoding distribution.
This allows our adaptive entropy bottleneck to adapt to the input distribution, rather than using a static dataset-optimized encoding distribution which suffers from the amortization gap.
We demonstrate that our low-cost input-adaptive method can be used by models containing the entropy bottleneck in order to achieve better rate-distortion performance.
Our code is available online%
\footnote{%
  \url{https://github.com/multimedialabsfu/learned-compression-of-encoding-distributions}%
}.

\section{Related works}
\label{sec:pdf_compression/related}

In their "scale hyperprior" architecture, Ballé~\emph{et~al.}~\cite{balle2018variational} utilize an entropy bottleneck to compress the latent representation $\boldvar{z} = h_a(\boldvar{y})$.
This is known as "side-information," and its compressed bitstream is transmitted before the bitstream for $\boldvar{y}$.
The reconstructed latent $\boldvar{\hat{z}}$ is then used to determine the parameters%
\footnote{\cite{balle2018variational} fixes $\boldvar{\mu} = 0$, though later papers~\cite{minnen2018joint} allow $\boldvar{\mu}$ to vary.}
$\boldvar{\mu}$ and $\boldvar{\sigma}$ of a Gaussian conditional distribution which is used to reconstruct $\boldvar{\hat{y}}$.
In many such models that utilize the scale hyperprior, the rate usage of the entropy bottleneck often comprises roughly 10\% of the total rate.
Evidently, the entropy bottleneck is a key component of many state-of-the-art (SOTA) image compression models, and so it is important to address the amortization gap suffered by the entropy bottleneck.


Balcilar~\emph{et~al.}~\cite{balcilar2022amortizationgap} propose non-learned methods for the compression and transmission of encoding distributions.
For the factorized entropy model, they first measure a discrete histogram of the $c$-th latent channel.
Then, they construct an approximation of the histogram using a $K$-Gaussian mixture model parameterized by $w^{(k)}$, $\mu^{(k)}$, and $\sigma^{(k)}$.
The aforementioned parameters are compactly encoded into 8~bits each, and transmitted.
That is, for a model using $C$ activated channels, $(3K - 1) \cdot C \cdot 8$ bits are required to transmit the parameters.
(Typically, $K \in [1, 3]$ is used, and $C \in [16, 192]$ for low-rate models.)
The approximation model is then used as the encoding distribution
$p_{\boldvar{\hat{y}}}(\boldvar{\hat{y}}_c) = \sum_{k=1}^K w_k \; \mathcal{\hat{N}}(\boldvar{\hat{y}}_c ; \, \mu_k, \, {\sigma_k}^2)$.
For the Gaussian conditional entropy model, they instead transmit a factor $\beta^{(c)}$ that corrects the error in the center (and most likely) bin's probability.
A key difference between our work and this prior work is that we formulate a learnable method for the compression of the encoding distribution.

Galpin~\emph{et~al.}~\cite{galpin2023entropy} introduce a lightweight non-learned context model for raster-scan ordered encoding, similar to Minnen~\emph{et~al.}~\cite{minnen2018joint}.
They compare their method in conjunction with~\cite{campos2019content,balcilar2022amortizationgap} in various configurations on both the standard factorized model and its GDN $\to$ ReLU replacement variant.
Our work is also broadly related to source instance overfitting for compression~\cite{vanRozendaal_overfitting_ICLR2021,Zhang_overfitting_VCIP2021}.
However, unlike these approaches, we do not overfit the neural network, but rather the latent distribution of the source instance, which is computationally more feasible.

%

\section{Proposed method}

\subsection{Compression of encoding distributions}



Let $\boldvar{x}$ be an input image, and $\boldvar{y} = g_a(\boldvar{x})$ be its transformed latent representation, and $\boldvar{\hat{y}} = \operatorname{Quantize}[\boldvar{y}]$ be its quantized version.
Consider a fully-factorized entropy model~\cite{balle2018variational}, which models the $j$-th channel of $\boldvar{\hat{y}}$ (denoted as $\boldvar{\hat{y}}_j$) using a single channel-specific encoding distribution $\boldvar{\hat{p}}_j$.
In this setting, all elements within a given channel are modelled using the same encoding distribution.
Then, the "true" distribution $\boldvar{p}_j$ can be determined exactly by computing the normalized histogram of $\boldvar{\hat{y}}_j$.
Notably, $\boldvar{p}_j$ is also the distribution with the minimum rate cost for encoding $\boldvar{\hat{y}}_j$ using the entropy model.
That is, the ideal encoding distribution $\boldvar{\hat{p}}^*_j$ is given by
\begin{equation*}
  \boldvar{\hat{p}}^*_j = \argmin_{\boldvar{\hat{p}}_j} \left[
    H(\boldvar{p}_j) + D_{\mathrm{KL}}(\boldvar{p}_j \parallel \boldvar{\hat{p}}_j)
  \right]
  \implies \boldvar{\hat{p}}^*_j = \boldvar{p}_j.
\end{equation*}
Unfortunately, using the true distribution $\boldvar{p}_j$ for encoding is infeasible, since the decoder does not have access to it.
Instead, we propose a method that generates a reconstructed approximation $\boldvar{\hat{p}}$ that targets $\boldvar{p}$, subject to a rate trade-off.

\begin{figure}[tbp]
  \centering
  \vspace{8pt}
  \includegraphics[width=\linewidth]{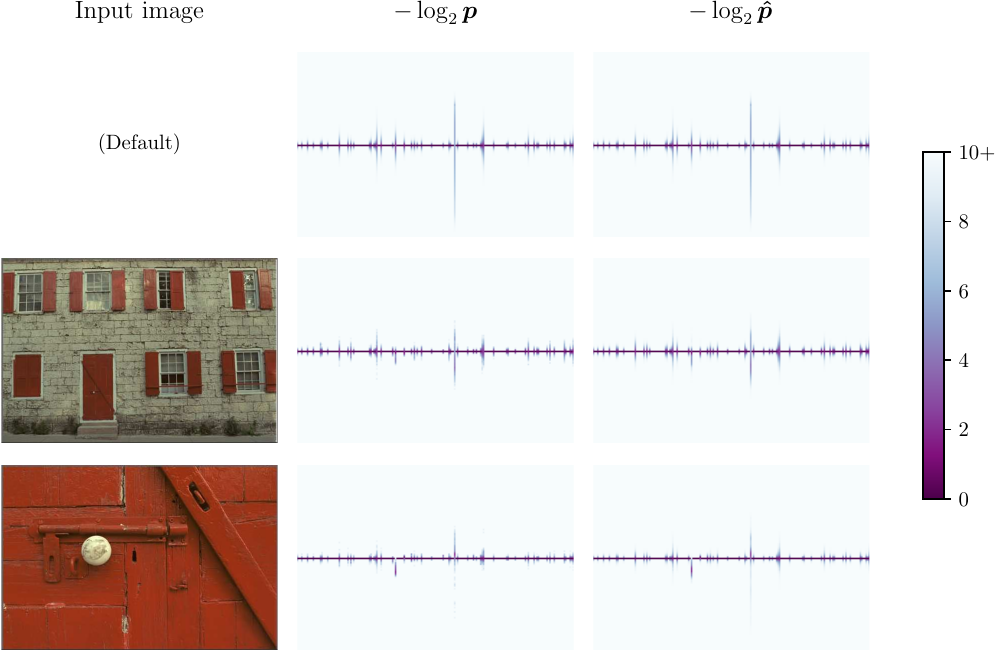}
  \caption[Visualization of target and reconstructed encoding distributions]{%
    Visualization of target ($\boldvar{p}$) and reconstructed ($\boldvar{\hat{p}}$) encoding distributions.
    Our proposed method reconstructs $\boldvar{\hat{p}}$, which is then used by the fully-factorized entropy model to encode the latent derived from a given input image.
    Each collection of distributions is visualized as a color plot,
    with channels varying along the $x$-axis,
    bins varying along the $y$-axis, and
    negative log-likelihoods represented by the $z$-axis (i.e., color).
  }
  \label{fig:pdf/pdfs}
\end{figure}

\cref{fig:pdf/pdfs} visualizes various collections of encoding distributions used for encoding different input images using a fully-factorized entropy model.
The left column shows various input images --- except for the "(Default)" image, which is an abstract amortized representation of randomly sampled images from the training dataset.
The middle and right columns visualize the negative log-likelihoods (in bits) of the true ($\boldvar{p}$) and reconstructed ($\boldvar{\hat{p}}$) encoding distributions, respectively.
Each collection of distributions is visualized as a color plot, where
\begin{enumerate}[label=(\roman*), noitemsep, topsep=0pt]
  \item the $x$-coordinate is the channel index $j$ of the latent $\boldvar{y}$,
  \item the $y$-coordinate is the bin index $i$ of the encoding distribution, and
  \item the $z$-coordinate (i.e., color intensity) is the negative log-likelihood of $(\boldvar{p}_j)_i$ or $(\boldvar{\hat{p}}_j)_i$, clipped within $[0, 10]$.
\end{enumerate}

As shown, the default amortized encoding distributions account for a wide range of possible input distributions.
However, the ideal encoding distributions for any given image often have much narrower ranges.
The average cost of encoding a randomly drawn sample from a distribution $\boldvar{p}$ is equal to the cross-entropy between that distribution and the one used for encoding, $\boldvar{\hat{p}}$.
Thus, a mismatch between the two leads to excessive rate usage.
In contrast, with our proposed method, the encoding distributions more closely approximate the ideal distributions for encoding a specific image.
Our method balances the trade-off between the rate cost of encoding the encoding distributions $\boldvar{\hat{p}}$ and the rate cost of encoding the latent representation $\boldvar{\hat{y}}$ using the encoding distributions.
Indeed, our method accurately reconstructs regions of high probability, which contribute the most towards saving rate.

\subsection{Architecture overview}
\label{sec:pdf_compression/architecture_overview}

\begin{figure}[htbp]
  \centering
  \vspace{-7pt}%
  \includegraphics[width=0.96\linewidth]{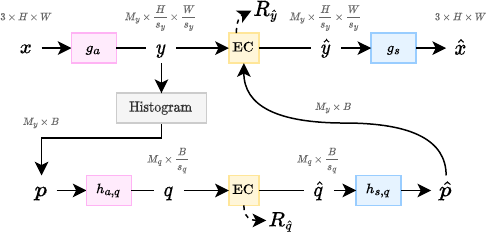}
  \vspace{-8pt}%
  \caption[Adaptive encoding distribution architecture]{%
    Adaptive encoding distribution architecture.%
  }
  \label{fig:pdf/arch}
\end{figure}

Our proposed method applied to the traditional "fully-factorized" architecture~\cite{balle2018variational} is visualized in \cref{fig:pdf/arch}.
In this configuration, the latent representation $\boldvar{y}$ is first passed through the histogram layer described in \cref{sec:pdf_compression/histogram}, which estimates the target distribution $\boldvar{p}$.
Then, $\boldvar{p}$ is passed through a distribution compression architecture consisting of the analysis transform $h_{a,q}$ which generates
a latent representation $\boldvar{q}$, followed by an entropy model (e.g., entropy bottleneck), and then finally a synthesis transform $h_{s,q}$ which reconstructs the distribution $\boldvar{\hat{p}}$.
The rate cost of encoding $\boldvar{q}$ is labelled $R_{q}$, and the rate cost of encoding $\boldvar{y}$ using $\boldvar{\hat{p}}$ is labelled $R_{y}$.

Like the scale hyperprior from~\cite{balle2018variational}, our proposed method also requires the transmission of side-information from which the encoding distributions are reconstructed.
However, the scale hyperprior applies 2D convolutions to downsample the 3D input tensor, and later reconstructs encoding distributions that vary along all 3 dimensions.
In contrast, our method initially collapses the spatial dimensions into a 1D histogram, to which it applies 1D convolutions, and later reconstructs encoding distributions that vary only along the channel dimension.
This allows our method to be more computationally efficient when the number of bins and encoding distributions is small.


\subsection{Histogram estimation}
\label{sec:pdf_compression/histogram}

To adapt the encoding distribution to the input, we must first determine the ideal target encoding distribution.
We do this by estimating the histogram for a given latent channel.
To ensure differentiability backwards through the histogram module, we use the method of kernel density estimation (KDE)~\cite{parzen1962estimation}.


Let $[y_{\mathrm{min}}, y_{\mathrm{max}}]$ be a supporting domain that encompasses all values $y_1, y_2, \ldots, y_{N}$ in a specific channel.
Let $\{b_1, b_2, \ldots, b_B\}$ denote the bin centers of the $B$-bin histogram, where $b_i = y_{\mathrm{min}} + i - 1$.
In the method of kernel density estimation, which is used to estimate the probability density function from a set of observations, a kernel function is placed centered at each observed sample, and those functions are summed to create the overall density function.
To construct the histogram, we evaluate the kernel density estimate at each $b_i$.
Or rather, we use an equivalent formulation where we instead place a kernel function at each bin center.
Then, the probability mass in the $i$-th bin can be estimated as
\begin{equation*}
  (\operatorname{Hist}(\boldvar{y}))_i =
  \frac{1}{\mathrm{Normalization}}
  \sum_{n=1}^{N} K\left(\frac{y_n - b_i}{\Delta b}\right),
\end{equation*}
where $\Delta b = b_{i+1} - b_i$ is the bin width, and $K(u)$ is the kernel function, which we choose to be the triangular function
\begin{equation*}
  K_{\mathrm{soft}}(u) = \max \{ 0, 1 - |u| \}.
\end{equation*}
Conveniently, the triangular kernel function ensures that the total mass of a single observation sums to $1$. 
Thus, to ensure that the total mass of the histogram under all $N$ observations sums to $1$, we may simply choose $\mathrm{Normalization} = N$.

While the above provides a piecewise differentiable estimate of the histogram, it is also possible to determine the exact \emph{hard} histogram by using a rectangular kernel function, which allocates $y_n$ to its nearest bin:
\begin{equation*}
  K_{\mathrm{hard}}(u) = \operatorname{rect}(u).
\end{equation*}
We can then use the "straight-through" estimator (STE)~\cite{bengio2013estimating} trick to get the benefits of both:
\begin{equation*}
  \operatorname{Hist}(\boldvar{y}) =
  \operatorname{Hist}_{\mathrm{soft}}(\boldvar{y}) +
  \operatorname{detach}(
    \operatorname{Hist}_{\mathrm{hard}}(\boldvar{y}) -
    \operatorname{Hist}_{\mathrm{soft}}(\boldvar{y})
  ).
\end{equation*}
Thus, the hard histogram is used for the forward pass, and the 
soft histogram for the backward pass (i.e., backpropagation).


\subsection{Loss function}
\label{sec:pdf_compression/loss}

Our model follows the variational formulation based on~\cite{balle2018variational}; further details on the construction may be found in earlier works such as~\cite{kingma2013autoencoding,balle2017endtoend,theis2017lossy}.
Briefly, for a variable $\boldvar{y}$, its counterparts $\tilde{\boldvar{y}}$ and $\hat{\boldvar{y}}$ denote the noise-modelled (for training) and quantized (for inference) versions of $\boldvar{y}$, respectively.
Furthermore, the variational encoder $q_{\boldvar{\phi}}(\boldvar{\tilde{y}}, \boldvar{\tilde{q}} \mid \boldvar{x})$ with trainable parameters $\phi$ is a probabilistic
%
model of the joint distribution for $\boldvar{\tilde{y}}$ and $\boldvar{\tilde{q}}$, for a given input $\boldvar{x}$.
Then, the loss function that we seek to minimize over the input data distribution $p_{\boldvar{x}}(\boldvar{x})$ is
\begin{equation*}
  \begin{split}
    \mathcal{L}
    &=
    \E_{\boldvar{x} \sim p_{\boldvar{x}}(\boldvar{x})}
    \E_{
      \boldvar{\tilde{y}}, \boldvar{\tilde{q}}
      \sim q_{\boldvar{\phi}}(\boldvar{\tilde{y}}, \boldvar{\tilde{q}} \mid \boldvar{x})
    }
    \left[
    \begin{multlined}
      -\log p_{\boldvar{\tilde{y}} \mid \boldvar{\tilde{q}}}(\boldvar{\tilde{y}} \mid \boldvar{\tilde{q}}) \\
      - \lambda_q \log p_{\boldvar{\tilde{q}}}(\boldvar{\tilde{q}}) \\
      + \lambda_x D(\boldvar{x}, \boldvar{\tilde{x}})
    \end{multlined}
    \right]
    \\
    &=
    \E_{\boldvar{x} \sim p_{\boldvar{x}}(\boldvar{x})}
    \left[
      R_{\boldvar{\tilde{y}}}(\boldvar{x})
      + \lambda_q R_{\boldvar{\tilde{q}}}(\boldvar{x})
      + \lambda_x D(\boldvar{x}, \boldvar{\tilde{x}})
    \right].
  \end{split}
\end{equation*}
For a given input $\boldvar{x}$, $R_{\boldvar{\tilde{y}}}(\boldvar{x})$ and $R_{\boldvar{\tilde{q}}}(\boldvar{x})$ are the rate costs of $\boldvar{\tilde{y}}$ and $\boldvar{\tilde{q}}$, respectively.
Furthermore, $D(\boldvar{x}, \boldvar{\tilde{x}})$ is a distortion metric --- typically mean-squared error (MSE) --- between the input $\boldvar{x}$ and its reconstruction $\boldvar{\tilde{x}}$.
Additionally, $\lambda_x$ is a trade-off hyperparameter between the rate and distortion.
And lastly, $\lambda_q$ is a trade-off hyperparameter between the rate cost of encoding $\boldvar{\tilde{q}}$ against the amount of rate saved when using $\boldvar{\tilde{q}}$ to encode $\boldvar{\tilde{y}}$.

The choice for $\lambda_q$ depends upon the ratio between the target and trained-upon input dimensions:
\begin{equation*}
  \lambda_q =
  \frac%
  {H_{\boldvar{x}, \mathrm{trained}} W_{\boldvar{x}, \mathrm{trained}}}%
  {H_{\boldvar{x}, \mathrm{target}} W_{\boldvar{x}, \mathrm{target}}}.
\end{equation*}
For instance, for the target dimension of $768 \times 512$ (i.e., images from the Kodak test dataset) and a trained-upon dimension of $256 \times 256$ (i.e., image patches from the Vimeo-90K dataset), we set $\lambda_q = \frac{1}{6}$.

Let $\boldvar{p}_j = (p_{j1}, \ldots, p_{jB})$ refer to the discrete $B$-bin probability distribution
used to encode the $j$-th channel of $\boldvar{\tilde{y}}$.
Then concretely, the rate cost of $\boldvar{\tilde{y}}$ is the cross-entropy between the true and reconstructed%
\footnote{Since the reconstructed distribution is determined directly from $\boldvar{y}$ rather than from $\boldvar{\tilde{y}}$, we denote it by $\boldvar{\hat{p}}$, not $\boldvar{\tilde{p}}$.}
discretized distributions,
\begin{equation*}
  R_{\boldvar{\tilde{y}}}
  =
  \sum_{j=1}^{M}
  \sum_{i=1}^{B}  
  -p_{ji} \log \hat{p}_{ji}.
\end{equation*}
Similarly, let $\boldvar{\tilde{q}}_j$ denote the $j$-th channel of $\boldvar{\tilde{q}}$.
Then, the rate cost of $\boldvar{\tilde{q}}$ is measured by the typical calculation,
\begin{equation*}
  R_{\boldvar{\tilde{q}}}
  =
  \sum_{j=1}^{M}
  \sum_{i=1}^{\operatorname{length}(\boldvar{l}_j)}
  -{l_{ji}} \log {l_{ji}},
  \quad \text{where } {\boldvar{l}_j} = p_{\boldvar{\tilde{q}}_j}(\boldvar{\tilde{q}}_j).
\end{equation*}

\subsection{Optimization}
\label{sec:pdf_compression/optimization}

We will now compute the derivatives for our proposed method.
To simplify the notation, 
we will focus on a single channel --- the $j$-th channel --- and directly denote $\boldvar{\tilde{y}}_j$ as $\boldvar{y}$, and $\boldvar{p}_j$ as $\boldvar{p}$.
From the perspective of the backward pass,
\begin{equation*}
  \begin{split}
    p_i
    = (\operatorname{Hist}(\boldvar{y}))_i
    &= \frac{s^2}{H W}
      \sum_n K_{\mathrm{soft}}{\left(\frac{y_n - b_i}{\Delta b}\right)} \\
    &= \frac{s^2}{H W}
      \sum_n \max \left\{ 0, 1 - \left| \frac{y_n - b_i}{\Delta b} \right| \right\},
  \end{split}
\end{equation*}
where $s$ is the downscale factor (e.g., $s = 2^4$),
and the dimension of the input image $\boldvar{x}$ is $H \times W$.
And so,
%
\begin{equation*}
  \resizebox{\linewidth}{!}{$
  \begin{aligned}
    \frac{\partial p_i}{\partial y_k}
    &= \frac{s^2}{H W}
      \sum_n \frac{\partial}{\partial y_k} \left[
        \max \left\{ 0, 1 - \left| \frac{y_n - b_i}{\Delta b} \right| \right\}
      \right] \\
    &= \frac{s^2}{H W}
      \frac{\partial}{\partial y_k} \left[
        \max \left\{ 0, 1 - \left| \frac{y_k - b_i}{\Delta b} \right| \right\}
      \right] \\
    &= \frac{s^2}{H W} 
      \frac{-1}{\Delta b} \cdot
      \underbrace{%
        H{\left( 1 - \left| \frac{y_k - b_i}{\Delta b} \right| \right)}
      }_{1 \text{ if } |y_k - b_i| < \Delta b \text{ else } 0}
      \cdot
      \underbrace{%
        \left( 2 \cdot H{\left( \frac{y_k - b_i}{\Delta b} \right)} - 1 \right)
      }_{-1 \text{ if } y_k < b_i \text{ else } 1}.
  \end{aligned}
  $}
\end{equation*}
%
%
The corresponding derivative for the reconstructed distribution $\boldvar{\hat{p}}$ may be determined as
\begin{equation*}
  \begin{split}
    \frac{\partial \hat{p}_i}{\partial y_k}
    &= \sum_l
      \frac{\partial \hat{p}_i}{\partial p_l}
      \frac{\partial p_l}{\partial y_k}
    \approx \sum_l
      \delta_{il}
      \frac{\partial p_l}{\partial y_k}
    = \frac{\partial p_i}{\partial y_k},
  \end{split}
\end{equation*}
where we have assumed $\frac{\partial \hat{p}_i}{\partial p_l} \approx \delta_{il}$, and that the dominant term in the sum is the term where $l = i$.

The rate cost of encoding the channel $\boldvar{y}$ with its associated encoding distribution $\boldvar{p}$ is given by
\begin{equation*}
  R_y = \frac{H W}{s^2} \sum_i -p_i \log \hat{p}_i.
\end{equation*}
Then, we may compute the derivative as follows:
%
\begin{equation}
  \label{eq:pdf_compression/optimization/dRdy_proposed}
  \resizebox{\linewidth}{!}{$
  \begin{aligned}
    \frac{\partial R_y}{\partial y_k}
    &= \frac{-H W}{s^2} \sum_i \frac{\partial}{\partial y_k} \left[ p_i \log \hat{p}_i \right]
    \\
    &= \frac{-H W}{s^2} \sum_i \left[
      \frac{p_i}{\hat{p}_i} \frac{\partial \hat{p}_i}{\partial y_k}
      + \frac{\partial p_i}{\partial y_k} \log \hat{p}_i
    \right]
    \\
    &\approx \frac{-H W}{s^2} \sum_i \left[
      \frac{p_i}{\hat{p}_i} \frac{\partial p_i}{\partial y_k}
      + \frac{\partial p_i}{\partial y_k} \log \hat{p}_i
    \right]
    \quad \text{assume } \frac{\partial \hat{p}_i}{\partial y_k} \approx \frac{\partial p_i}{\partial y_k}
    \\
    &= \frac{-H W}{s^2} \sum_i
      \left[ \frac{p_i}{\hat{p}_i} + \log \hat{p}_i \right]
      \frac{\partial p_i}{\partial y_k}
    \\
    %
    %
    %
    &= \frac{-HW}{s^2}
      \sum_{\scriptscriptstyle i \in \{
          \lfloor{y_k - y_{\mathrm{min}}}\rfloor + 1,
          \lceil{y_k - y_{\mathrm{min}}}\rceil + 1
      \}}
      \left[ \frac{p_i}{\hat{p}_i} + \log \hat{p}_i \right]
      \frac{\partial p_i}{\partial y_k}
    \\
    %
    &= \frac{-1}{\Delta b}
      \left(
      \begin{multlined}
        \left[ {\frac{p_i}{\hat{p}_i} + \log \hat{p}_i} \right]_{
          i=\lceil{y_k - y_{\mathrm{min}}}\rceil + 1
        } \\
        \qquad
        -
        \left[ {\frac{p_i}{\hat{p}_i} + \log \hat{p}_i} \right]_{
          i=\lfloor{y_k - y_{\mathrm{min}}}\rfloor + 1
        }
      \end{multlined}
      \right)
    \\
    &\approx \frac{-1}{\Delta b} \left(
      \log \hat{p}_{\scriptscriptstyle \lceil{y_k - y_{\mathrm{min}}}\rceil + 1} -
      \log \hat{p}_{\scriptscriptstyle \lfloor{y_k - y_{\mathrm{min}}}\rfloor + 1}
    \right)
    \enspace
    \text{assume } \hat{p}_i \approx p_i.
  \end{aligned}
  $}
\end{equation}
%
Thus, assuming that $\hat{p}_i \approx p_i$ and $\frac{\partial \hat{p}_i}{\partial y_k} \approx \frac{\partial p_i}{\partial y_k}$, the gradient of the rate cost $R_y$ is directly proportional to the difference in code lengths of the two bins whose centers are nearest to the value $y_k$.
Intuitively, this makes sense, since the rate cost of encoding the $k$-th element is directly proportional to the linear interpolation between the code lengths of the two nearest bins:
\begin{equation*}
  \begin{split}
    R_{y_k} =
    - \alpha \cdot
      \log \hat{p}_{\lfloor{y_k - y_{\mathrm{min}}}\rfloor + 1}
    - (1 - \alpha) \cdot
      \log \hat{p}_{\lceil{y_k - y_{\mathrm{min}}}\rceil + 1},
  \end{split}
\end{equation*}
where $\alpha
= (y_k - b_{\lfloor{y_k - y_{\mathrm{min}}}\rfloor + 1}) / \Delta b$.
The linear interpolation may be justified by the fact that $\hat{y}_k$ inhabits only a single bin at a time.
The probability of $\hat{y}_k$ inhabiting the left bin is $\alpha$, and the probability of inhabiting the right bin is $1 - \alpha$.
This aligns with Shannon's measure of entropy, which is the expected value of the code length.


For comparison, the standard "entropy bottleneck" represents the likelihood of a symbol by
\begin{equation*}
  p(y)
  = c{\left(y + \sfrac{1}{2}\right)} - c{\left(y - \sfrac{1}{2}\right)}
  = \int_{y - \sfrac{1}{2}}^{y + \sfrac{1}{2}} f(t) \, dt,
\end{equation*}
where $c$ is a cumulative distribution function and $f(y) = \frac{d}{dy} c(y)$ is a probability density function, from which the encoding distribution can be determined by integrating over the bin.
Then, the rate cost for the "entropy bottleneck" is merely the sum of the negative log-likelihoods (i.e., code lengths),
\begin{equation*}
  R_y = \sum_i -\log p(y_i).
\end{equation*}
We may then compute its derivative as follows:
\begin{equation}
  \label{eq:pdf_compression/optimization/dRdy_standard}
  \begin{split}
    \frac{\partial R_y}{\partial y_k}
    &= - \sum_i \frac{\partial}{\partial y_k} \left[ \log p(y_i) \right]
    = - \frac{\partial}{\partial y_k} \left[ \log p(y_k) \right] \\
    &= - [p(y_k)]^{-1} \cdot \left[
      f{\left(y_k + \sfrac{1}{2}\right)} - f{\left(y_k - \sfrac{1}{2}\right)}
    \right].
  \end{split}
\end{equation}

Interestingly, whereas the derivative for the proposed method (under the assumption that $\hat{p}_i \approx p_i$) computed in \cref{eq:pdf_compression/optimization/dRdy_proposed} contains the difference between the "right" and "left" log-likelihoods,
the derivative for the entropy bottleneck computed in \cref{eq:pdf_compression/optimization/dRdy_standard} contains the difference between the "right" and "left" evaluations of the probability density function.

\section{Experimental setup}
\label{sec:pdf_compression/experimental_setup}

\subsection{Architecture details}
\label{sec:pdf_compression/experimental_setup/architecture_details}

\begin{figure}[htbp]
  \centering
  \includegraphics[width=0.85\linewidth]{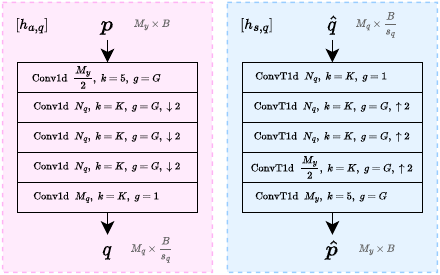}
  \caption[Architecture layer diagram for $h_{a,q}$ and $h_{s,q}$ transforms]{%
    Architecture layer diagram for $h_{a,q}$ and $h_{s,q}$ transforms.
    $k$ denotes kernel size, $g$ denotes number of channel groups, and $\downarrow, \uparrow$ denote stride.%
  }
  \label{fig:pdf/arch-hasq}
\end{figure}

As shown in \cref{fig:pdf/arch-hasq}, our $h_{a,q}$ and $h_{s,q}$ are implemented using a simple five-layer convolutional neural network.
In between each of the convolutional layers shown is a ReLU activation function, as well as a channel shuffle operation, as is done in ShuffleNet~\cite{zhang2017shufflenet}.
There are two strides of length $2$, resulting in a total downscaling factor of $s_q = 4$.
For all our models, we set $K = 15$ to control the kernel sizes, and $G = 8$ to control the number of channel groups.
Furthermore, we set $(N_q, M_q) = (32, 16)$ for low-resolution models, and $(N_q, M_q) = (64, 32)$ for high-resolution models.
The number of bins $B$ is set between $128$ and $1024$ across different models.
We have elected to use an entropy bottleneck design for simplicity, though one can likely further improve the compression performance of the encoding distribution compression architecture by using more powerful entropy modeling techniques (e.g., a scale hyperprior).
Since $\lambda_q$ is a parameter that depends on the ratio between the target and trained-upon input dimensions, it may also be advisable to train a single model that supports $\lambda$-rate control strategies (e.g., G-VAE~\cite{cui2020gvae,cui2022asymmetric} and QVRF~\cite{tong2023qvrf}) for both latent representations $y, q$.
However, we have not yet explored this possibility.

\subsection{Training details}
\label{sec:pdf_compression/experimental_setup/training_details}

Our models are trained on $256 \times 256$ image patches from the Vimeo-90K triplet dataset~\cite{xue2019video}.
A training batch size of $16$ was used, along with the Adam optimizer~\cite{kingma2014adam} with an initial learning rate of $10^{-4}$ that was decayed by a factor of $0.1$ whenever the validation loss plateaued.
Specifically, we loaded the weights of the pretrained models from CompressAI~\cite{begaint2020compressai}, which were also trained using the same setup as above.
We replaced the static distributions of the \texttt{EntropyBottleneck} module with the dynamically generated adaptive distributions from our proposed encoding distribution compression module.
Then, we froze the weights for $g_a$ and $g_s$, and trained only the weights for our encoding distribution compression model (i.e., for $h_{a,q}$ and $h_{s,q}$, and the entropy model for $q$).
Finally, we evaluated our models on the standard Kodak test dataset~\cite{kodak_dataset} containing 24 images of size $768 \times 512$.

\section{Experimental results}
\label{sec:pdf_compression/experimental_results}

\begin{figure}[bp]
  \centering
  \includegraphics[width=\linewidth]{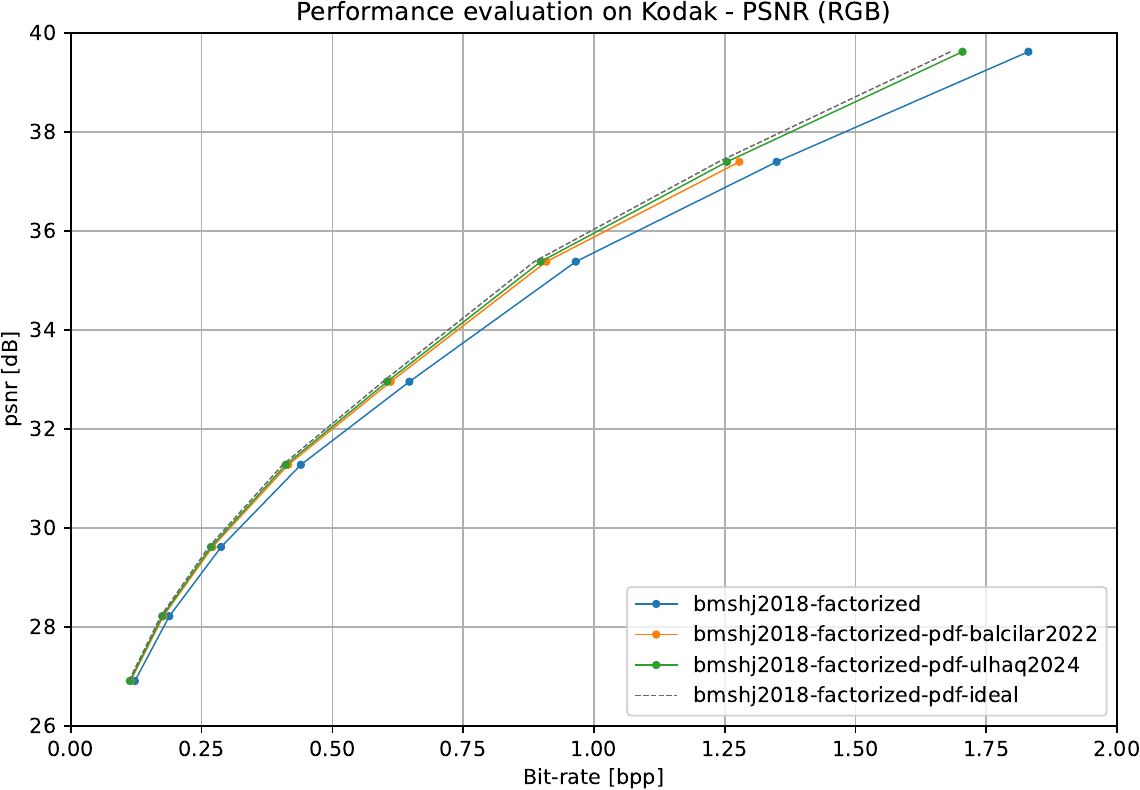}
  \caption[RD curves for Kodak dataset]{%
    RD curves for the Kodak dataset.%
  }
  \label{fig:pdf/rd-curves}
\end{figure}

\cref{fig:pdf/rd-curves} shows the rate-distortion (RD) curves comparing a given base model against the same model enhanced with an encoding distribution compression method.

%

In~\cref{tbl:pdf}, we compare the rate savings of our proposed method against the base model at various quality levels.
Additionally, we also list the maximum rate savings that can be theoretically achieved by perfectly eliminating the amortization gap between the true and reconstructed distributions, i.e., by using the true distribution directly as the encoding distribution with zero additional transmission cost.
As shown, our approach achieves a 7.10\% BD-rate reduction over the base model, in comparison to a maximum possible reduction of 8.21\% BD-rate reduction achievable by perfectly eliminating the amortization gap.

\begin{table}[htbp]
  \centering
  \vspace{\tablecaptionaboveskip}%
  \caption[Rate savings for bmshj2018-factorized per-quality]{%
    Potential and achieved rate savings for the bmshj2018-factorized model~\cite{balle2018variational} when equipped with our proposed distribution compression method.%
  }
  \vspace{\tablecaptionbelowskip}%
  \label{tbl:pdf}
  \renewcommand{\arraystretch}{0.9}
  \setlength{\tabcolsep}{4.25pt}
  \renewcommand\theadfont{\scriptsize}
  \footnotesize
  \begin{tabular}[]{ccccccc}
    \toprule
    %
    \thead{Quality}
    & \thead{Original \\ (bpp)}
    & \thead{Potential \\ gain (bpp)}
    & \thead{Potential \\ gain (\%)}
    & \thead{Our \\ (bpp)}
    & \thead{Our gain \\ (bpp)}
    & \thead{Our gain \\ (\%)} \\
    \midrule
    1 & 0.122 & -0.012 & -9.45 & 0.113 & -0.010 & -8.06 \\
    2 & 0.188 & -0.016 & -8.66 & 0.174 & -0.014 & -7.53 \\
    3 & 0.287 & -0.024 & -8.20 & 0.267 & -0.020 & -7.13 \\
    4 & 0.440 & -0.035 & -7.90 & 0.409 & -0.031 & -6.95 \\
    5 & 0.647 & -0.051 & -7.83 & 0.602 & -0.045 & -6.98 \\
    6 & 0.965 & -0.080 & -8.25 & 0.898 & -0.067 & -6.95 \\
    7 & 1.349 & -0.111 & -8.24 & 1.254 & -0.095 & -7.04 \\
    8 & 1.830 & -0.149 & -8.14 & 1.705 & -0.126 & -6.88 \\
    \midrule
    * &       &        & -8.21 &       &        & -7.10 \\
    \bottomrule
  \end{tabular}
\end{table}

In order to compute the potential savings, we first ran the base model on each image from the Kodak test dataset~\cite{kodak_dataset}, giving us a collection of true distributions $\boldvar{p}$ for each image.
Furthermore, the base model provides a static encoding distribution $\boldvar{p_{\textnormal{default}}}$.
Then, we computed the potential savings (in bpp) for each image as
$(\Delta R)_{\textnormal{max}}
= \frac{1}{s^2}
\sum_j D_{\mathrm{KL}}(\boldvar{p}_j \parallel (\boldvar{p_{\textnormal{default}}})_j)$.
We then averaged these values across the dataset.

In \cref{tbl:rate-gains}, we compare the rate savings for various base models equipped with a distribution compression method.
"Ratio" is the fraction of the total rate occupied by the original unmodified model.
"Gap" is the maximum potential rate gain with perfect distribution reconstruction and zero additional transmission cost.
"Gain" is the actual rate gain achieved by the model.
All models are evaluated on the Kodak dataset~\cite{kodak_dataset}.

\begin{table}[htbp]
  \centering
  \vspace{\tablecaptionaboveskip}%
  \caption[Comparison of rate savings for various models]{%
    Comparison of rate savings for various models.%
  }
  \vspace{\tablecaptionbelowskip}%
  \label{tbl:rate-gains}
  \renewcommand{\arraystretch}{0.9}
  \setlength{\tabcolsep}{0.75pt}
  \renewcommand\theadfont{\scriptsize}
  \footnotesize
  \begin{tabular}[]{cccccc}
    \toprule
    \multirow{4}{*}{\thead{Model}}  
    & \multirow{4}{*}{\thead{Quality}\hspace{2pt}}
    & \multicolumn{3}{c}{\thead{Factorized}}
    & \multicolumn{1}{c}{\thead{Total}}
    \\
    \cmidrule(lr){3-5}
    \cmidrule(lr){6-6}
    &
    & \thead{Ratio \\ (\%)}
    & \thead{Gap   \\ (\%)}
    & \thead{Gain  \\ (\%)}
    & \thead{Gain  \\ (\%)}
    \\
    \midrule
    bmshj2018-factorized~\cite{balle2018variational} + Balcilar2022~\cite{balcilar2022amortizationgap} & 1
      &   100 & -9.45 & -6.79 & -6.79 \\  
    bmshj2018-factorized~\cite{balle2018variational} + ours & 1
      &   100 & -9.45 & -8.06 & -8.06  \\
    bmshj2018-factorized~\cite{balle2018variational} + Balcilar2022~\cite{balcilar2022amortizationgap} & *
      &   100 & -8.21 & -5.74 & -5.74  \\  
    bmshj2018-factorized~\cite{balle2018variational} + ours & *
      &   100 & -8.21 & -7.10 & -7.10  \\  
    \bottomrule
  \end{tabular}
\end{table}

\cref{tbl:pdf/params-macs} reports the number of trainable parameters and the number of multiply-accumulate operations (MACs) per pixel for various model configurations.
The results are calculated assuming an input image size of $768 \times 512$, and a latent representation size of $48 \times 32$.
As shown, our model requires far fewer parameters and MACs/pixel than the comparable scale hyperprior~\cite{balle2018variational} model.
In particular, for comparable configurations, our method's $h_{a,q}$ and $h_{s,q}$ transforms require $96\text{--}97\%$ fewer parameters and $96\text{--}99\%$ fewer MACs/pixel than the scale hyperprior's $h_{a}$ and $h_{s}$ transforms.

\begin{table}[htbp]
  \centering
  \vspace{\tablecaptionaboveskip}%
  \caption[Trainable parameters and MACs per pixel]{%
    Trainable parameter counts and number of multiply-accumulate operations (MACs) per pixel.
  }
  \vspace{\tablecaptionbelowskip}%
  \label{tbl:pdf/params-macs}
  \renewcommand{\arraystretch}{0.9}
  \setlength{\tabcolsep}{1.75pt}
  \renewcommand\theadfont{\scriptsize}
  \footnotesize
  \begin{tabular}[]{ccccc}
    \toprule
    \thead{Model configuration}
    & \thead{Params}
    & \thead{MAC/px}
    & \thead{Params}
    & \thead{MAC/px}
    \\
    \midrule
    \thead{$(M_y, N_q, M_q, K, G, B)$}
    & \multicolumn{2}{c}{\thead{$h_{a,q}$}}
    & \multicolumn{2}{c}{\thead{$h_{s,q}$}}
    \\
    \cmidrule(lr){1-1}
    \cmidrule(lr){2-3}
    \cmidrule(lr){4-5}
    %
    %
    %
    %
    %
    Ours $(192, 32, 16, 15, 8, 256)$  & 0.029M & 10 & 0.029M & 10 \\
    Ours $(320, 64, 32, 15, 8, 1024)$ & 0.097M & 126 & 0.097M & 126 \\[3pt]
    \midrule
    \thead{$(N, M)$}
    & \multicolumn{2}{c}{\thead{$h_{a}$}}
    & \multicolumn{2}{c}{\thead{$h_{s}$}}
    \\
    \cmidrule(lr){1-1}
    \cmidrule(lr){2-3}
    \cmidrule(lr){4-5}
    bmshj2018-hyperprior~\cite{balle2018variational} $(128, 192)$ & 1.040M & 1364 & 1.040M & 1364 \\
    bmshj2018-hyperprior~\cite{balle2018variational} $(192, 320)$ & 2.396M & 3285 & 2.396M & 3285 \\[3pt]
    \bottomrule
  \end{tabular}
\end{table}

\section{Conclusion}
\label{sec:pdf_compression/conclusion}

In this paper, we proposed a learned method for the compression of encoding distributions.
Our method effectively measures and compresses the encoding distributions used by the entropy bottleneck.
The experiments we performed where we only trained the distribution compression component show that this method is effective at significantly reducing the amortization gap.
Since many learned compression models use the entropy bottleneck component, our method provides them with a low-cost improvement in bitrate.
Furthermore, our work opens up the possibility of using learned distribution compression as a paradigm for correcting encoding distributions.


\subsection{Future work}
\label{sec:pdf_compression/conclusion/future_work}

A few steps remain towards making encoding distribution corrective methods viable parts of more advanced entropy models such as~\cite{balle2018variational,cheng2020learned,he2022elic}.
These include the following:
\begin{itemize}
  \item
    The proposed adaptive entropy bottleneck needs to be formulated in such a way that it can be trained fully end-to-end.
    When training a pretrained base model equipped with our adaptive distribution compression module, we found that unfreezing the pretrained transform weights led to a degradation in RD performance.
  \item
    Application of our adaptive distribution method to the Gaussian conditional component of entropy models.
    This component (along with the entropy bottleneck) is used to construct entropy modeling methods such as those used in~\cite{balle2018variational,cheng2020learned,he2022elic}.
    Most SOTA learned image compression models predict the location and scale parameters of the Gaussian 
    encoding distributions. 
    There are potential rate improvements to be made by adapting the shapes of the encoding distributions to those that more closely match the data distribution.
    %
\end{itemize}

\phantomsection

\bibliographystyle{IEEEbib}
\bibliography{references}

\end{document}